\pdfoutput=1
\RequirePackage{ifpdf}

\documentclass[cits,11pt,a4paper]{article}
\usepackage{jinstpub}
\usepackage{float}

\usepackage{caption}
\usepackage{subcaption}

\usepackage{graphicx}
\usepackage{natbib}
\usepackage{lineno}
\title{Demonstration of Event Position Reconstruction based on Diffusion in the NEXT-White Detector}

\author[1]{J.~Haefner,}
\author[2,a]{K.E.~Navarro\note[a]{Corresponding author},}
\author[3]{R.~Guenette,}
\author[2]{B.J.P.~Jones,}
\author[2]{A.~Tripathi,}
\author[4]{C.~Adams,}
\author[3]{H.~Almaz\'an,}
\author[5]{V.~\'Alvarez,}
\author[6]{B.~Aparicio,}
\author[7]{A.I.~Aranburu,}
% ...
\affiliation[1]{
Department of Physics, Harvard University, Cambridge, MA 02138, USA}
\affiliation[2]{
Department of Physics, University of Texas at Arlington, Arlington, TX 76019, USA}
\affiliation[3]{
Department of Physics and Astronomy, Manchester University, Manchester. M13 9PL, United Kingdom}
\affiliation[4]{
Argonne National Laboratory, Argonne, IL 60439, USA}
\affiliation[5]{
Instituto de Instrumentaci\'on para Imagen Molecular (I3M), Centro Mixto CSIC - Universitat Polit\`ecnica de Val\`encia, Camino de Vera s/n, Valencia, E-46022, Spain}
\affiliation[6]{
Department of Organic Chemistry I, University of the Basque Country (UPV/EHU), Centro de Innovaci\'on en Qu\'imica Avanzada (ORFEO-CINQA), San Sebasti\'an / Donostia, E-20018, Spain}
\affiliation[7]{
Department of Applied Chemistry, Universidad del Pais Vasco (UPV/EHU), Manuel de Lardizabal 3, San Sebasti\'an / Donostia, E-20018, Spain}
% ...

\abstract{Noble element time projection chambers are a leading technology for rare event detection in physics, such as for dark matter and neutrinoless double beta decay searches. Time projection chambers typically assign event position in the drift direction using the relative timing of prompt scintillation and delayed charge collection signals, allowing for reconstruction of an absolute position in the drift direction. In this paper,  alternate methods for assigning event drift distance via quantification of electron diffusion in a pure high pressure xenon gas time projection chamber are explored. Data from the NEXT-White detector demonstrate the ability to achieve good position assignment accuracy for both high- and low-energy events. Using point-like energy deposits from $^{83\mathrm{m}}$Kr calibration electron captures ($E\sim45$~keV), the position of origin of low-energy events is determined to $2~$cm precision with bias $< 1~$mm. A convolutional neural network approach is then used to quantify diffusion for longer tracks (E$\geq$~1.5~MeV), yielding a precision of 3~cm on the event barycenter. The precision achieved with these methods indicates the feasibility energy calibrations of better than 1\%~FWHM at Q$_{\beta\beta}$ in pure xenon, as well as the potential for event fiducialization in large future detectors using an alternate method that does not rely on primary scintillation.}

\begin{document}
\maketitle

\section{Introduction}

Noble element time projection chambers (TPCs) in the liquid or gaseous phase are a widely used technology for rare event searches. These include the NEXT \cite{next2022}, EXO/nEXO \cite{exo2019,nEXO2021}, and PandaX \cite{pandax2021} experiments for neutrinoless double beta decay ($0 \nu \beta \beta$) searches, and the XENON \cite{aprile2017xenon1t}, LUX-ZEPLIN \cite{lz2020}, and DarkSide \cite{DarkSide2018} experiments for dark matter searches, among others. The basic operating principle of the TPC is that when a particle interacts in the detector, it produces a flash of light through primary scintillation (S1), and ionization electrons along the path of the particle. Using uniform electric fields applied across the detector volume, the ionization electrons are drifted with a known velocity and collected by a readout system. In electroluminescent TPCs such as NEXT, charge is detected by driving the ionization electrons across a high voltage gap, called an electroluminescence region, in order to produce an amplified secondary scintillation signal (S2). For NEXT-White, the total drift length is 664.5~mm and the drift velocity is 0.91 mm/$\mu s$. The time difference between the S1 and S2 signals allows the determination of the position in the drift direction $z$, given a known drift velocity.  Thus full event reconstruction including absolute placement in $z$ requires both S1 and S2 signals to be employed.  

Information about $z$ is in principle accessible through other means than the S1-S2 time difference alone.  As the electron swarm is drifted under the applied electric field, it spreads with a width proportional to $\sqrt{z}$ due to diffusion. This results in pulses for the recorded S2 signal which are wider in time for events that have drifted from larger $z$. Consequently, the study of the signal shapes in the S2 pulse can in principle also be used to determine the $z$ position of an event. The NEXT program has characterized diffusion in xenon gas at various pressures and electric fields~\cite{simon2018electron,mcdonald2019electron}. At the 41~V/cm/bar operating point of NEXT-White, the longitudinal reduced diffusion constant is approximately $D_L=1000~\sqrt{\mathrm{bar}}~\mu \mathrm{m}/\sqrt{\mathrm{cm}}$ and the transverse reduced diffusion constant $D_T=3800~\sqrt{\mathrm{bar}}~\mu \mathrm{m}/\sqrt{\mathrm{cm}}$.

If achievable, this technique yields several advantages. One is that having redundant methodologies for determining event position can enable more cross checks, better position reconstruction, improved background rejection or selection efficiencies. For example, if a prospective event with matched S1 and S2 signals, is found to have an S2 width that is different than would be expected from diffusion given the time difference between the S1 and S2 signals, it can be rejected as having an incorrectly assigned S1, potentially through accidental coincidence. Furthermore, if a single S2 event is found accompanied by two potential S1 signals, it would traditionally be rejected. By using the diffusion information, the correct S1 signal can be identified, increasing the selection efficiency.  This is likely to be an especially useful technique for $^{83\mathrm{m}}$Kr calibration of large-scale future detectors~\cite{adams2021sensitivity}, where pileup of events could otherwise become a limiting factor in detector calibration and hence energy resolution. Finally, this method could also allow for a xenon TPC to operate with limited access (or even without) to S1 information. Although noble element TPCs have proven highly scalable to date, advancing to new detector scales  will present technical challenges. The light collection requirements for the small S1 signals are more severe than those for the larger S2 signals, the latter being amplified through electroluminescence. With multiple R\&D pathways now being explored to realize future very large xenon TPCs~\cite{avasthi2021kiloton,byrnes2023next,mcdonald2018demonstration,rivilla2020fluorescent,villalpando2020improving}, understanding the information content in each signal component is of significant interest.

In this paper, methods for identifying the position of an event in the drift (z) direction based on the signal width of the diffusion of the ionization electrons are demonstrated using the NEXT-White experiment. In Section~\ref{sec:new}, the detector and data set are briefly described. In Section~\ref{sec:kr}, two methods using signal width from diffusion in order to determine the $z$ position of point-like $^{83\mathrm{m}}$Kr calibration events are developed, which employ analytical quantification of the shape of the waveform. In Section~\ref{sec:he}, a method is described for extracting the $z$ position from events at higher energies where the more complex track typologies requires analysis via machine learning algorithms. In both cases, reconstruction of event $z$ position with few-cm precision is demonstrated. Finally, Section~\ref{sec:conc} presents the conclusions.

\section{The NEXT-White and NEXT-100 detectors}
\label{sec:new}

NEXT (Neutrino Experiment with a Xenon TPC) is an experimental program aiming at the detection of $0 \nu \beta \beta$ decay in $^{136}$Xe, using successive generations of high pressure gaseous xenon electroluminescent time projection chambers (HPXe EL-TPCs) \cite{Nygren:2009zz}. Small scale prototypes demonstrated the capability of the technology to achieve sub-$1\%$ FWHM energy resolution and to topologically identify signal-like events \cite{NEXT:Demo2013, osti_1271011}, and this capability has since been tested underground with the larger ($\sim$5~kg of $^{136}$Xe at 10~bar) NEXT-White detector \cite{NEXT:backgrounds2019, NEXT:ecal2019, NEXT:eventid2019, NEXT:NEW2018}, at the Laboratorio Subterráneo de Canfranc (LSC) in Spain.  In addition to measuring two-neutrino~\cite{novella2022measurement} and  demonstrating neutrinoless~\cite{novella2023demonstration} double beta decay searches based on event-by-event topological identification and a direct background subtraction between enriched and depleted xenon, NEXT-White has served as a test-bed to inform the designs of future NEXT experiments including NEXT-100~\cite{alvarez2012next} and ton-scale phases~\cite{adams2021sensitivity}.

\begin{figure}[htp!]
\centering
\includegraphics[width=0.7\linewidth]{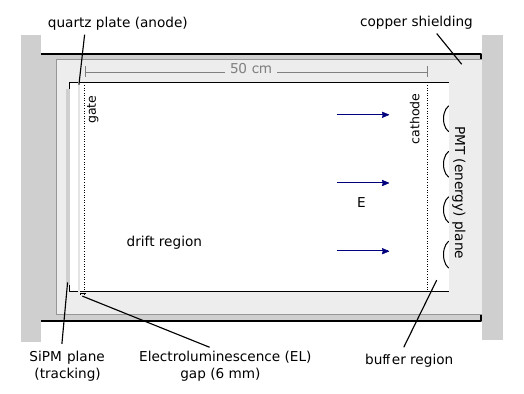}
\caption{Schematic of the EL-based TPC developed by the NEXT collaboration for neutrinoless double beta decay searches in $^{136}$Xe, from \cite{NEXT:ecal2019}.}
\label{fig:NEXT_TPC}
\end{figure}

The cylindrical NEXT-White TPC (shown schematically in Fig.~\ref{fig:NEXT_TPC}) has a length of 53~cm and a diameter of 40~cm. The energy of each event is measured by twelve Hamamatsu R11410-10 photomultiplier tubes (PMTs) placed $130~$mm from a transparent wire array cathode. The events are imaged by a 2D-array ($10~$mm pitch) of $1792$ SensL C-Series, $1~$mm$^2$ silicon photomultipliers (SiPMs), placed a few mm behind an electroluminescence (EL) gap of $6~$mm. The drift region has an electric field of $40~$V cm$^{-1}$ bar$^{-1}$ and the EL region is defined by a stainless steel mesh and a grounded quartz plate coated with indium tin oxide (ITO) and tetraphenyl butadiene (TPB) thin films. More details on the NEXT-White detector can be found in Ref.~\cite{NEXT:NEW2018}.

The  combination of the tracking information with the time of the event from the S1 signal (t$_0$) provides the 3D ($x$, $y$, $z$) positions of events. This information is typically needed for fiduacialization, to veto the edges of the detector where events are more likely to be background, and to apply the position dependent corrections for electron attachment required to achieve the target energy resolutions of $\sim$1\% FWHM~\cite{NEXT:ecal2019}. For NEXT-White, continuous detector calibration and monitoring was carried out by flowing radioactive $^{83\mathrm{m}}$Kr into the detector. $^{83\mathrm{m}}$Kr is a noble gas which decays via low-energy ($41.5~$keV) electron captures producing point-like events uniformly throughout the detector volume. This calibration allows to correct on a day-by-day basis for spatial variations in the detector, and for the finite electron lifetime caused by ionization electrons attaching to impurities before collection \cite{nextkr}.  

The coming phase of the NEXT program, NEXT-100 is presently under construction, and aims to demonstrate an ultra-low background  search for $0\nu\beta\beta$ in high pressure xenon gas at the 100~kg scale~\cite{martin2016sensitivity}.  The NEXT-100 TPC is approximately 1~m long and 1~m in diameter, scaling up linear dimensions of NEXT-White by a factor of two. 

\section{Reconstruction of the $z$ position of low-energy $^{83\mathrm{m}}$Kr electron captures using diffusion}
\label{sec:kr}

$^{83\mathrm{m}}$Kr has proven central to achieving position-dependent calibration of the NEXT detector, both for nonuniformities in ($x$, $y$) (the plane perpendicular to the drift direction), and for variations in $z$ (the drift direction) due to electron attachment. $^{83\mathrm{m}}$Kr decay events are excellent candidates to study the position reconstruction from diffusion, as they are close to point sources at production. This means that their width after diffusion can be straightforwardly quantified from the shape of the detected electron cloud. The S1 signals produced by $^{83\mathrm{m}}$Kr events are the lowest energy signals used in NEXT,  and their detection could thus be among the more challenging aspects of future large detector design.

The ionization cloud diffuses in both the transverse and the longitudinal directions during drift. A Gaussian electron cloud with longitudinal width $d$ traveling at velocity $v$ will produce an approximately Gaussian pulse of light with width in time of approximately $d / v$ when entering the EL gap, with a small correction from the time it takes to cross the gap. Non-Gaussian corrections to the pulse shape were studied in Ref.~\cite{mcdonald2019electron} and found to be negligible.  In contrast, the transverse width impacts the distribution of light across the SiPMs of the tracking plane, and its precision is limited by the 1~cm SiPM spacing. For this reason, the optimal diffusion-based measure of $z$ for krypton events is extracted from longitudinal diffusion only. The width of the pulse in time, also referred to as the ``(longitudinal) event width'', is measured in terms of the root mean squared (RMS) of the pulse.  According to the diffusion equation, the RMS$^2$ is expected to increase linearly with drift distance ($z$ position). Longitudinal diffusion in the NEXT-White detector has been previously quantified to have an RMS spread of $0.3\, \textup{mm} / \sqrt{\textup{cm}}$  \cite{NEXT:2018edrift}.

\begin{figure}[htb!]
\centering
  \centering
  \includegraphics[width=.49\linewidth]{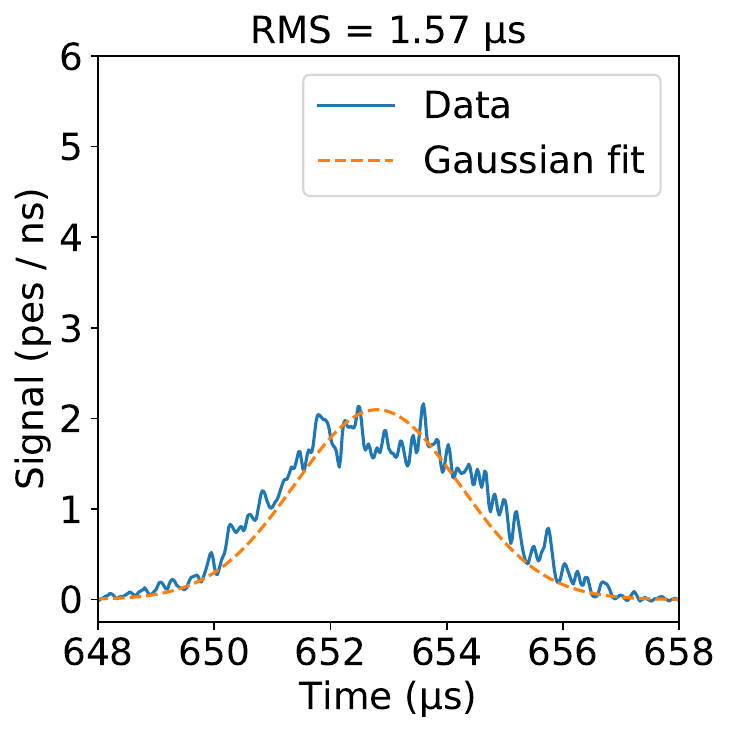} % was 92755
  \centering
  \includegraphics[width=.49\linewidth]{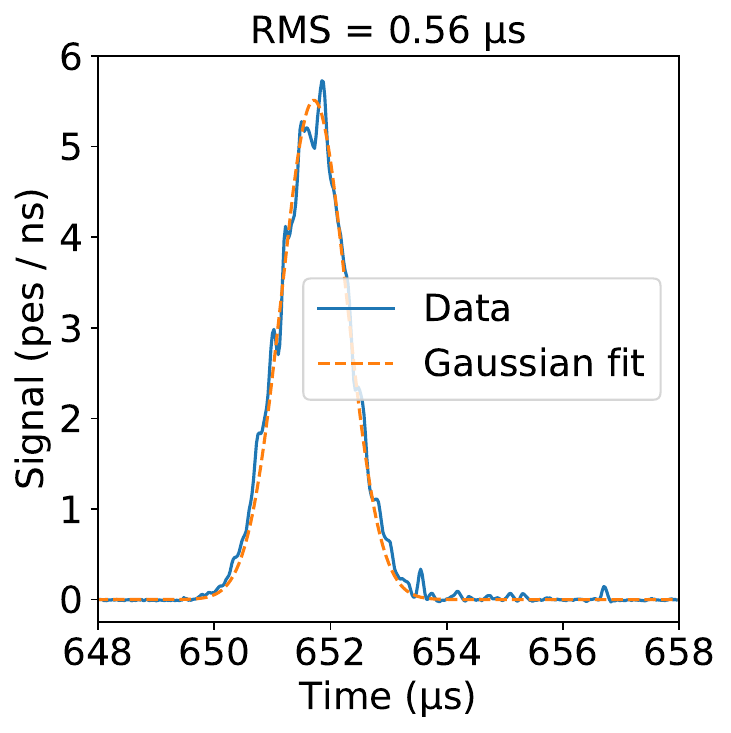} % was 92763
\centering
\caption[Example $^{83\mathrm{m}}$Kr events with widths.]{Two examples of $^{83\mathrm{m}}$Kr events as a function of time, where signals from all 12 PMTs are summed, overlaid with Gaussian fit with width fixed to calculated RMS value. Time widths of events as measured by root mean squared indicated above the corresponding plots.}
\label{fig:kr_examples}
\end{figure}

\begin{figure}[htb!]
\centering
\includegraphics[width=0.8\linewidth]{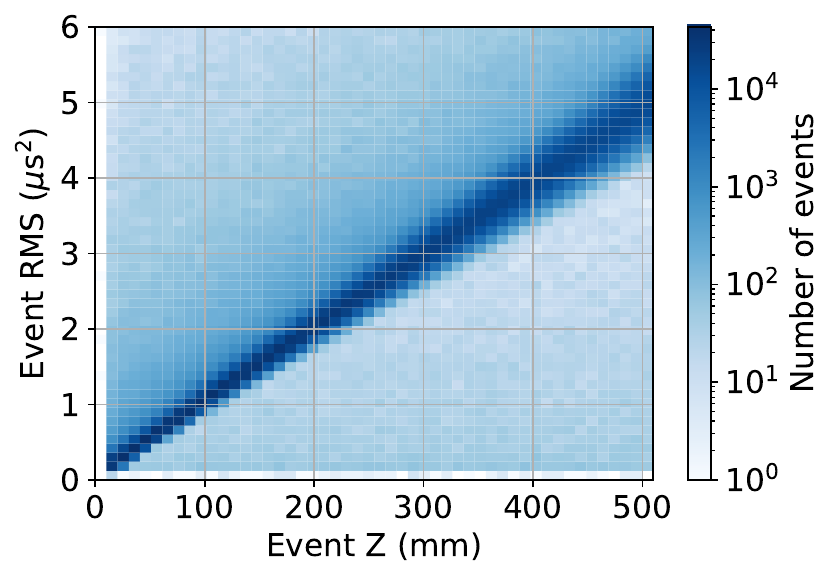}
\caption{The square of the longitudinal root mean squared (RMS$^2$) width of $^{83\mathrm{m}}$Kr events as a function of the $z$ position in the NEXT-White detector. A clear linear relationship between the two is observed, as expected.}
\label{fig:new_rms_v_z}
\end{figure}

The study presented here uses $7~$million $^{83\mathrm{m}}$Kr events taken over the course of a single day in NEXT-White. Two examples of $^{83\mathrm{m}}$Kr event pulses as a function of time can be seen in Fig.~\ref{fig:kr_examples}. In Fig.~\ref{fig:new_rms_v_z}, the distribution of event widths (in RMS$^2$) as a function of $z$ position (determined from S1) is shown. The linear increase of RMS$^2$ with $z$, as anticipated from diffusion, can be observed.

\begin{figure}[htb!]
\centering
    \begin{subfigure}[b]{0.49\textwidth}
    \centering
    \includegraphics[width=\textwidth]{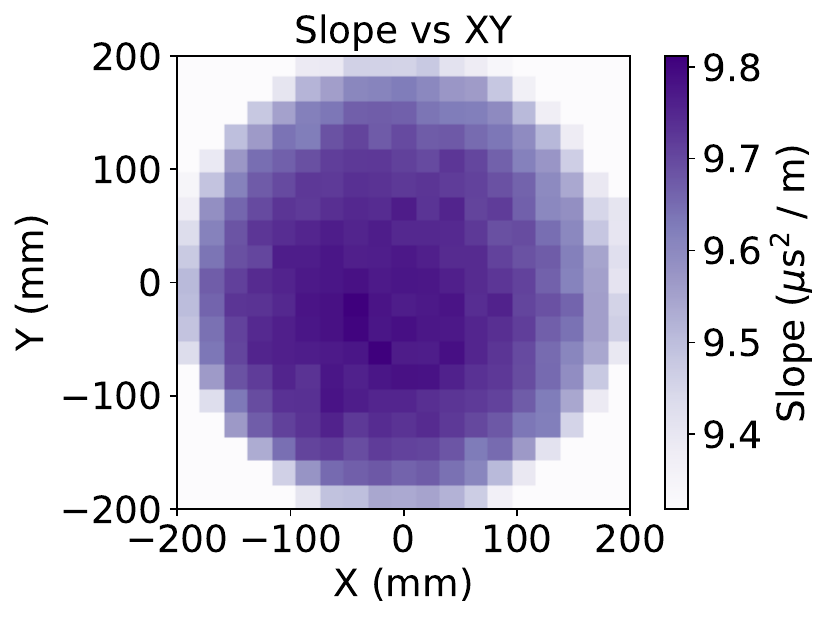}
\end{subfigure}
\hfill
    \begin{subfigure}[b]{0.49\textwidth}
    \centering
    \includegraphics[width=\textwidth]{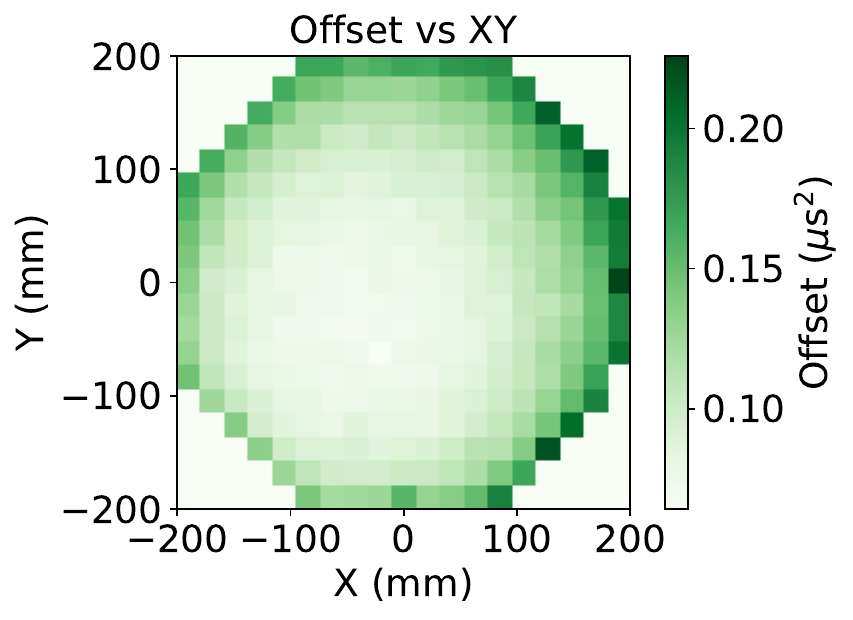}
\end{subfigure}
\caption{Linear fit parameters of $^{83\mathrm{m}}$Kr event RMS$^2$ as a function of $z$ (from S1) in NEXT-White for different $x$ and $y$ locations. Left: Slope of the linear fit, corresponding to diffusion. Right: Offset of the linear fit, corresponding to typical width of $^{83\mathrm{m}}$Kr event at $z$ $= 0$~mm. A clear dependence of both parameters with $x$ and $y$ is seen.}
\label{fig:new_params_v_$XY$}
\end{figure}

This measured linear relationship allows extraction of the $z$ position of a $^{83\mathrm{m}}$Kr event given the RMS$^2$ of the pulse, named $z_{\mathrm{RMS}}$. The offset corresponds to the width of a typical $^{83\mathrm{m}}$Kr event which occurs exactly at the EL gap, where there is almost no diffusion, while the slope corresponds to the impact of diffusion along $z$. These parameters are extracted as a function of $x$ and $y$ position. To provide $z_{\mathrm{RMS}}$ positions for all ($x$,$y$) locations, the geometry is sub-divided into $19 \times 19$ ($x$, $y$) bins, each $10.5 \times 10.5~$mm$^2$. For each bin, a linear fit to the relationship between RMS$^2$ and $z$ is performed and the values of slope and offset are extracted. The observed spatial variation of the fitted parameters is shown in Fig.~\ref{fig:new_params_v_$XY$}.  A plausible explanation for the small variations in the offset parameter are position dependence in the width of the EL gap. Variations in the fitted slope appear to reflect differences in the extracted  diffusion coefficient.  This could be a consequence of non-uniformity in the electric fields near the detector boundary.  These variations are small, with a standard deviation of 1.3\%. 

\begin{figure}[htb!]
\centering
    \begin{subfigure}[b]{0.48\textwidth}
    \centering
    \includegraphics[width=\textwidth]{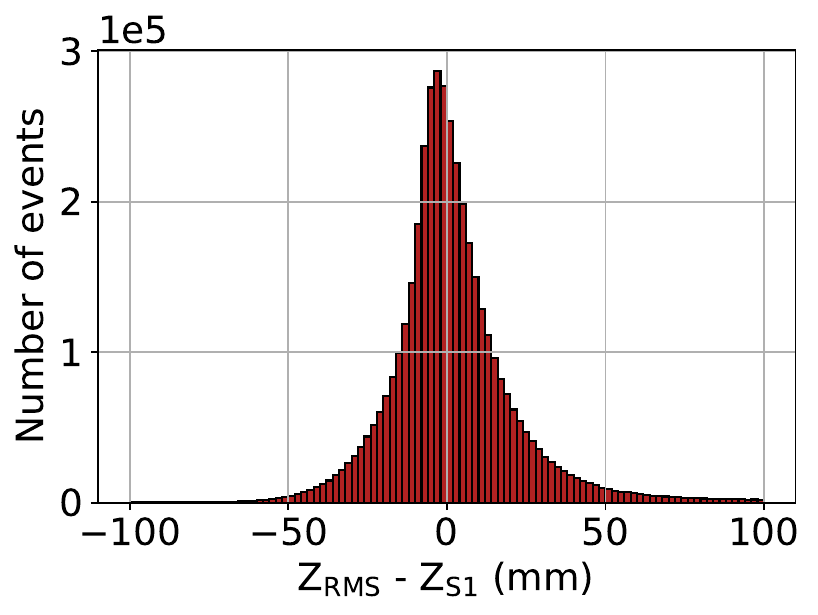}
\end{subfigure}
\hfill
    \begin{subfigure}[b]{0.50\textwidth}
    \centering
    \includegraphics[width=\textwidth]{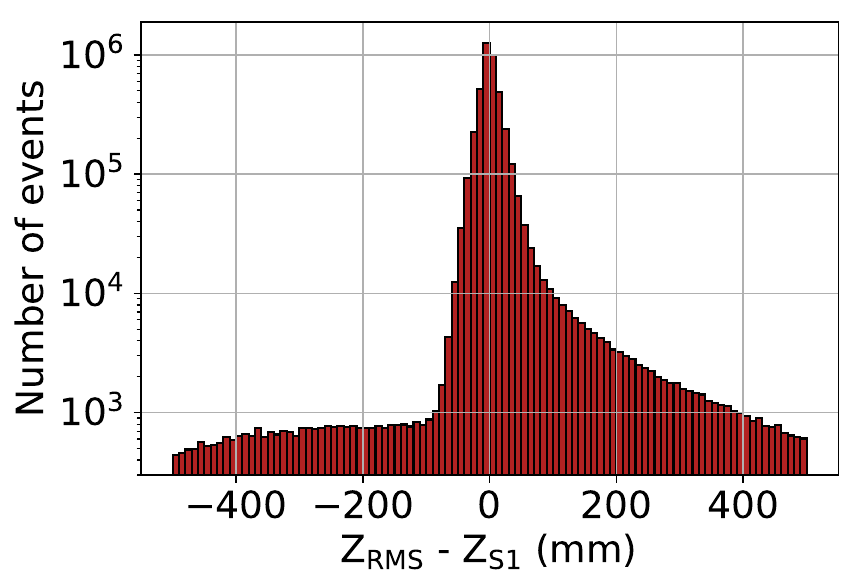}
\end{subfigure}
\caption{Differences between $z$ positions determined by RMS ($z_{\mathrm{RMS}}$) and determined from S1 ($z_{\mathrm{S1}}$) for $^{83\mathrm{m}}$Kr events in NEXT-White, shown in linear (left) and log (right) scales.}
\label{fig:new_dz}
\end{figure}

\begin{figure}[htb!]
\centering
    \begin{subfigure}[b]{0.497\textwidth}
    \centering
    \includegraphics[width=\textwidth]{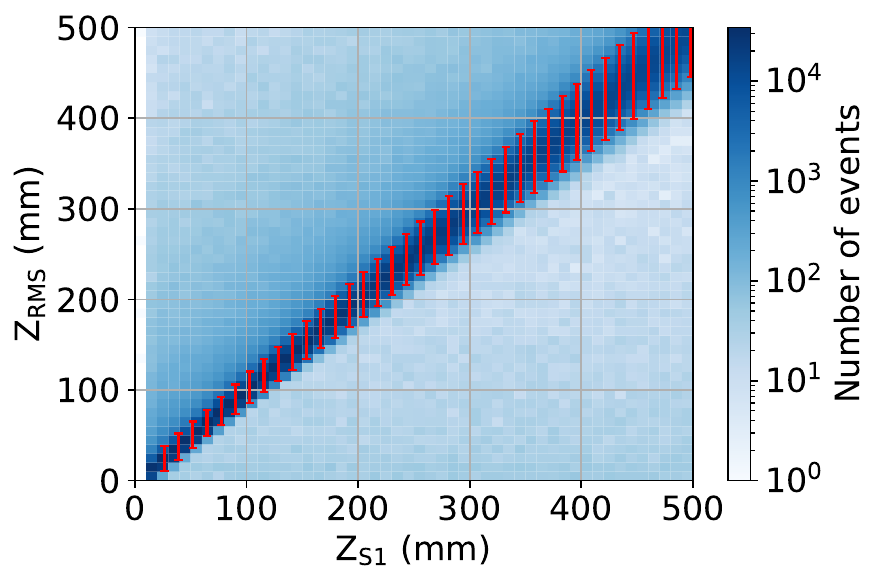}
\end{subfigure}
\hfill
    \begin{subfigure}[b]{0.497\textwidth}
    \centering
    \includegraphics[width=\textwidth]{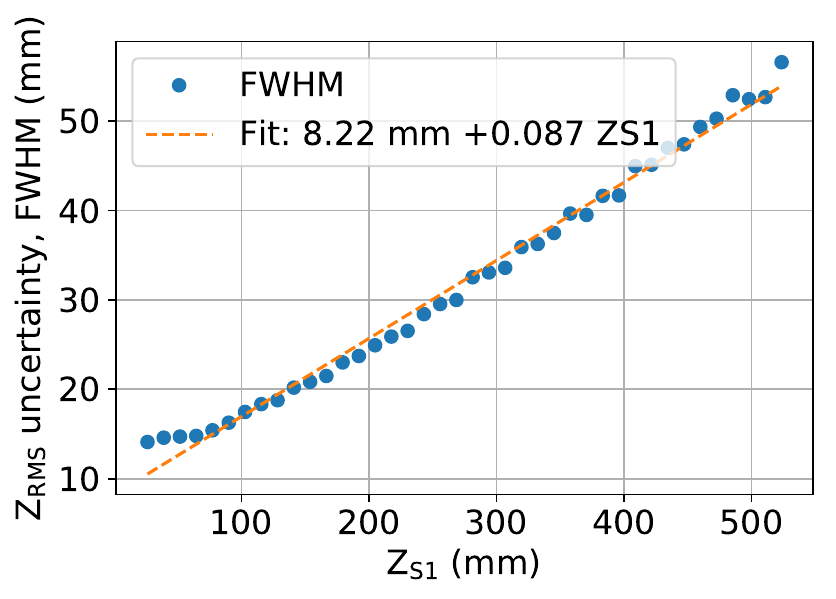}
\end{subfigure}
\caption{Left: $z$ position estimated from width ($z_{\mathrm{RMS}}$) in function of the $z$ position assigned from S1 ($z_{\mathrm{S1}}$) for $^{83\mathrm{m}}$Kr events in NEXT-White. Red uncertainties, representing FWHM of $z_{\mathrm{RMS}}$ in a given $z_{\mathrm{S1}}$ range, are overlaid. Right: FWHM of $z_{\mathrm{RMS}}$ as a function of $z_{\mathrm{S1}}$ in NEXT-White, with a linear fit, understood as the increase in uncertainty of the $z_{\mathrm{RMS}}$ with $z_{\mathrm{S1}}$.}
\label{fig:new_fwhm}
\end{figure}

The $z_{\mathrm{RMS}}$ position values obtained from the linear fit to the RMS$^2$ distributions as described above can be compared to the $z$ positions obtained from the S1 signal ($z_{\mathrm{S1}}$) in Fig.~\ref{fig:new_dz}. $z_{\mathrm{RMS}}$ is seen to have a small overall bias compared to $z_{\mathrm{S1}}$, with an overall median shift of $z_{\mathrm{RMS}}-z_{\mathrm{S1}}=-0.20$~mm. The error $|z_{\mathrm{RMS}} - z_{\mathrm{S1}}|$ averaged over the whole detector is 9.4~mm, indicating most events are estimated using the RMS method as within 1~cm of the position assigned using the S1 signal. Long and non-Gaussian tails on the positive end of the distribution of $z_{\mathrm{RMS}}-z_{\mathrm{S1}}$ indicate a population of events much wider (RMS much larger) than would be predicted from S1. This could be due to events with incorrectly assigned S1 pulses, for example in a case where part of the S2 signal is misinterpreted as an S1. The distribution of $z_{\mathrm{RMS}}$ as a function of $z_{\mathrm{S1}}$ is shown in Fig.~\ref{fig:new_fwhm}. The distribution is overlaid with error bars indicating the FWHM spread in the distribution of $z_{\mathrm{RMS}}$ values in fixed $z_{\mathrm{S1}}$ bins. These indicate the spread in assigned $z_{\mathrm{RMS}}$ values given a fixed, known $z_{\mathrm{S1}}$, and are interpreted as the uncertainty in the extraction of $z_{\mathrm{RMS}}$. The uncertainty is seen to increase linearly with $z_{\mathrm{S1}}$ at a rate of $89 \, \textup{mm}/\textup{m}$, as estimated from the right panel of Fig.~\ref{fig:new_fwhm}.

\begin{figure}[htp!]
\centering
    \begin{subfigure}[b]{0.485\textwidth}
    \centering
    \includegraphics[width=\textwidth]{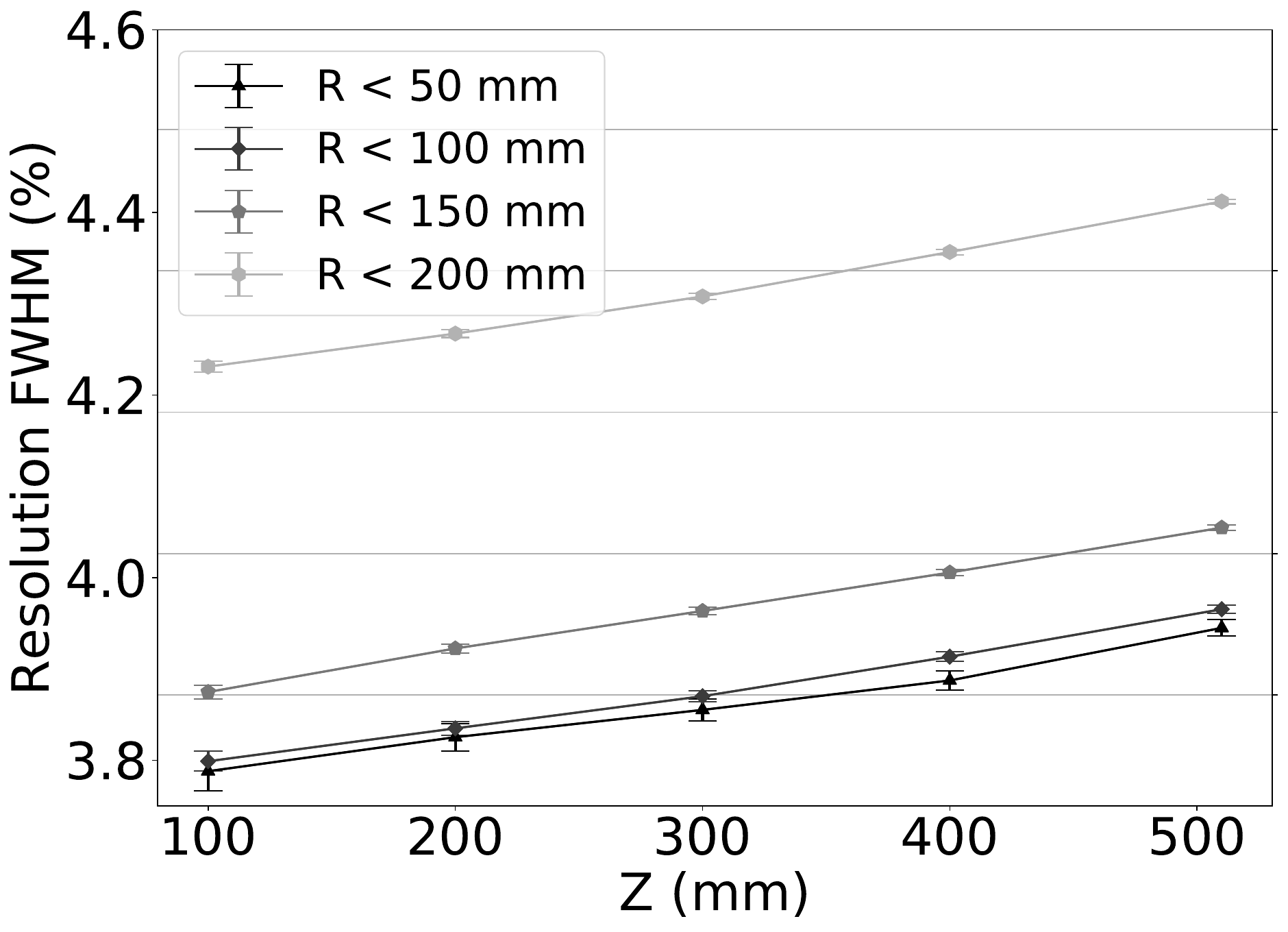}
\end{subfigure}
\hfill
    \begin{subfigure}[b]{0.495\textwidth}
    \centering
    \includegraphics[width=\textwidth]{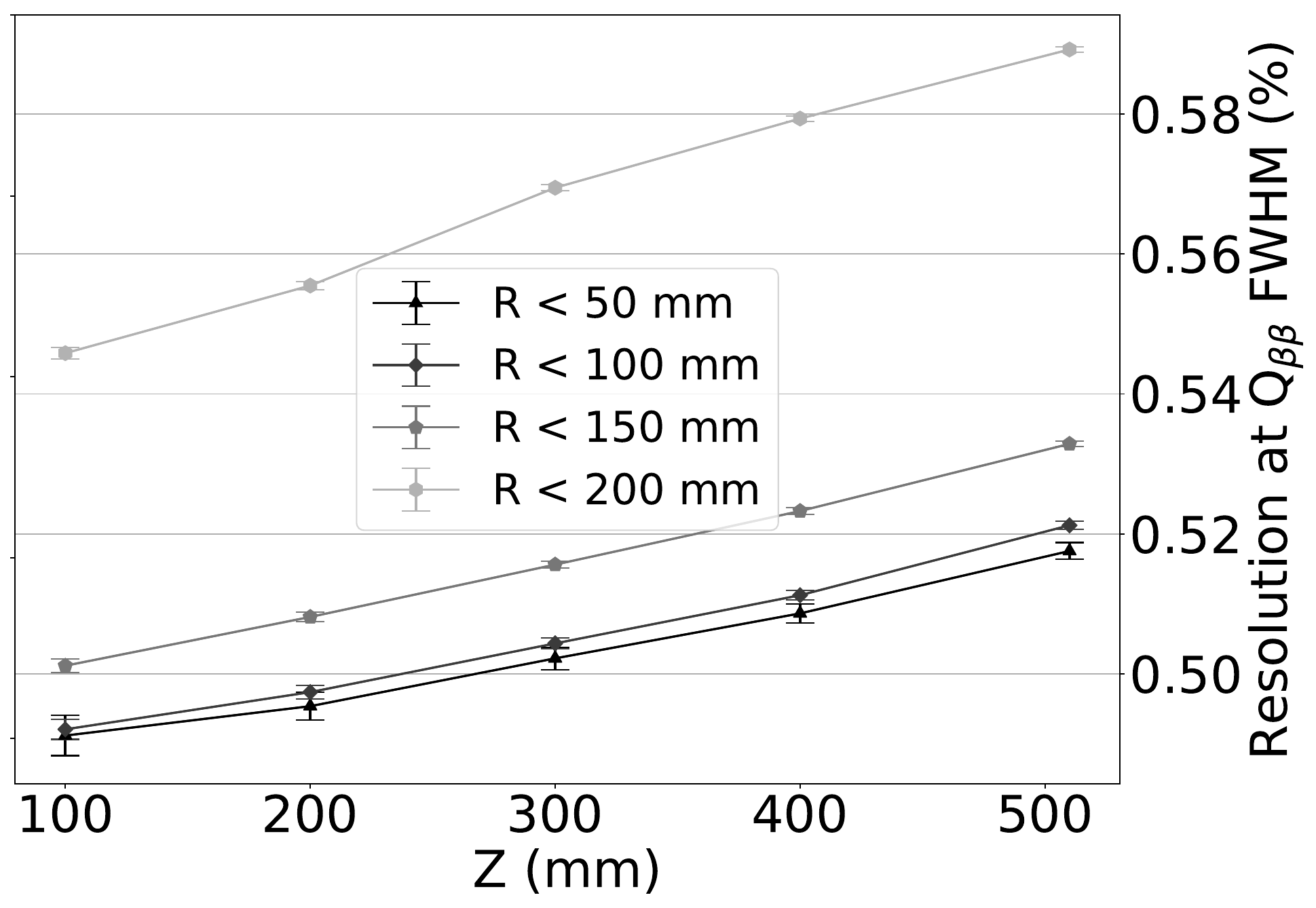}
\end{subfigure}
\caption{Energy resolution for $^{83\mathrm{m}}$Kr events as a function of $z_{\mathrm{S1}}$ in NEXT-White, for regions of varying maximum distance from central axis $R$. Left axis indicates resolution in FWHM / $41.5~$keV for both data sets, and right axis is matched to left axis to indicate resolution extrapolated to Q$_{\beta\beta}$ for both data sets. Volumes are overlapping, with $z_{\mathrm{S1}} = 300~$mm including all points with $z \leq 300~$mm, for example. Left: Energy resolution calculated using $z_{\mathrm{S1}}$. Right: Energy resolution calculated using z$_{\mathrm{RMS}}$.}
\label{fig:new_res}
\end{figure}

One of the key goals of using  $^{83\mathrm{m}}$Kr in NEXT is energy resolution calibration. Variability in the detected brightness of $^{83\mathrm{m}}$Kr events over the detector is used to generate the detector response correction that is applied to higher energy events.  Any imprecision in the $z$ reconstruction thus implies an imprecision in energy calibration.   
Energy resolution for $^{83\mathrm{m}}$Kr events is defined as the energy peak percent FWHM, and is measured for both $z_{\mathrm{RMS}}$ and $z_{\mathrm{S1}}$ as a function of position by subdividing the detector into several (overlapping) volumes of increasing maximum event radius ($r$) and $z$-position. The resolution comparison can be seen in Fig.~\ref{fig:new_res} for NEXT-White data, where a slight degradation in resolution is observed for $z_{\mathrm{RMS}}$ as compared to $z_{\mathrm{S1}}$.  This extrapolates to a change by around $0.01\%$ at $Q_{\beta \beta}$. Such a difference is sure to be negligible when determining  sensitivity to neutrinoless double beta decay.

%\subsection{Expected performance of the diffusion method in NEXT-100}

In order to analyze the applicability of the aforementioned method to larger detectors, 1 million $^{83\mathrm{m}}$Kr events were generated using a Monte Carlo simulation in the NEXT-100 detector, in a configuration resembling as close as possible the anticipated running configuration of the detector.

\begin{figure}[t]

\centering
    \begin{subfigure}[b]{0.50\textwidth}
    \centering
    \includegraphics[width=\textwidth]{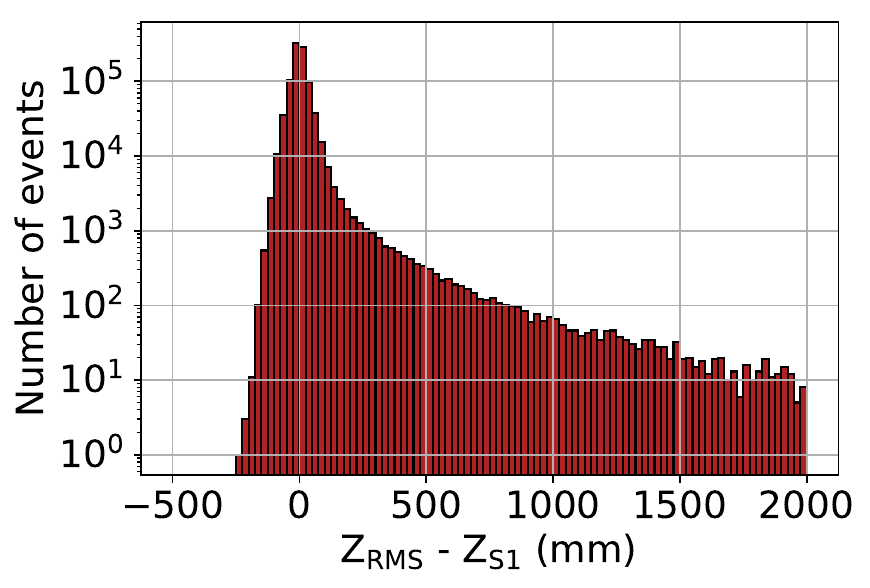}
\end{subfigure}
    \hfill
    \begin{subfigure}[b]{0.49\textwidth}
     \centering
     \includegraphics[width=\textwidth]{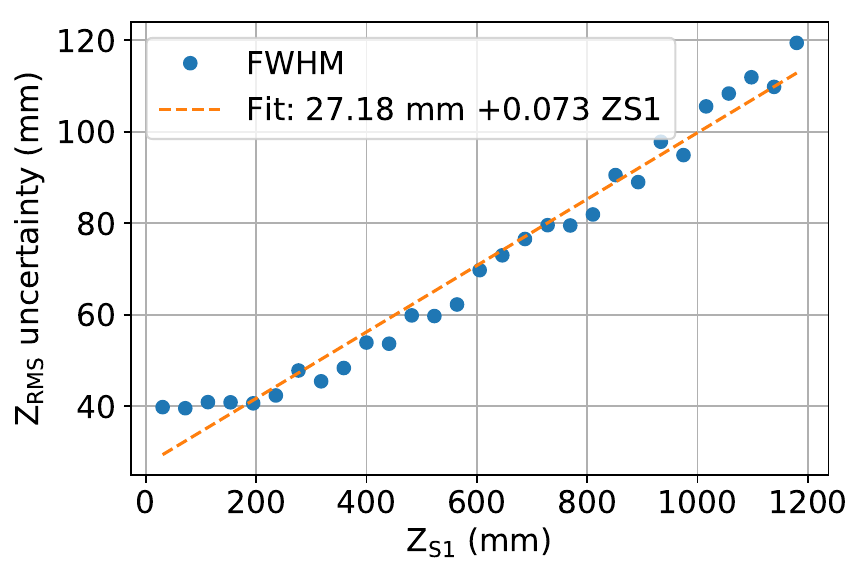}
 \end{subfigure}
\caption{NEXT-100 simulation. Left: Differences between $z$ position determined by RMS ($z_{\mathrm{RMS}}$) and determined from S1 ($z_{\mathrm{S1}}$) for $^{83\mathrm{m}}$Kr events shown in log scale. Right: FWHM of $z_{\mathrm{RMS}}$ as a function of $z_{\mathrm{S1}}$ with a linear fit, understood as the increase in uncertainty of the $z_{\mathrm{RMS}}$ with $z_{\mathrm{S1}}$.}
\label{fig:n100_newfig}
\end{figure}

The distribution of differences between the $z$ positions assigned from diffusion ($z_{\mathrm{RMS}}$), and from S1 ($z_{\mathrm{S1}}$) in NEXT-100 is shown in the left of Fig.~\ref{fig:n100_newfig}. The long non-Gaussian tails are comparable to those observed in NEXT-White, with median difference of -0.28~mm (compared to -0.20~mm in NEXT-White), again indicating a lack of significant bias in a particular direction. The error $|z_{\mathrm{RMS}} - z_{\mathrm{S1}}|$ averaged over the whole detector is 16.1~mm, somewhat larger than in NEXT-White.  The energy resolution obtained with z$_{\mathrm{RMS}}$ is around $0.01\%$ worse in each volume than that achievable with z$_{\mathrm{S1}}$, comparable to what was seen for NEXT-White. The increasing uncertainty as a function of $z$ thus translates to only a minuscule degradation of the energy resolution at the Q$_{\beta\beta}$ value. A similar linear relationship between uncertainty of $z_{\mathrm{RMS}}$ assignment as a function of $z_{\mathrm{S1}}$ can be seen in the right of Fig.~\ref{fig:n100_newfig}, although the slope predicted from Monte Carlo of NEXT-100 is somewhat lower than that observed in NEXT-White ($73$ mm/m, compared to $89$ mm/m in NEXT-White), indicating the more idealized simulation performs somewhat better than the real detector. 

\begin{figure}[htb!]
\centering
  \centering
  \includegraphics[width=.48\linewidth]{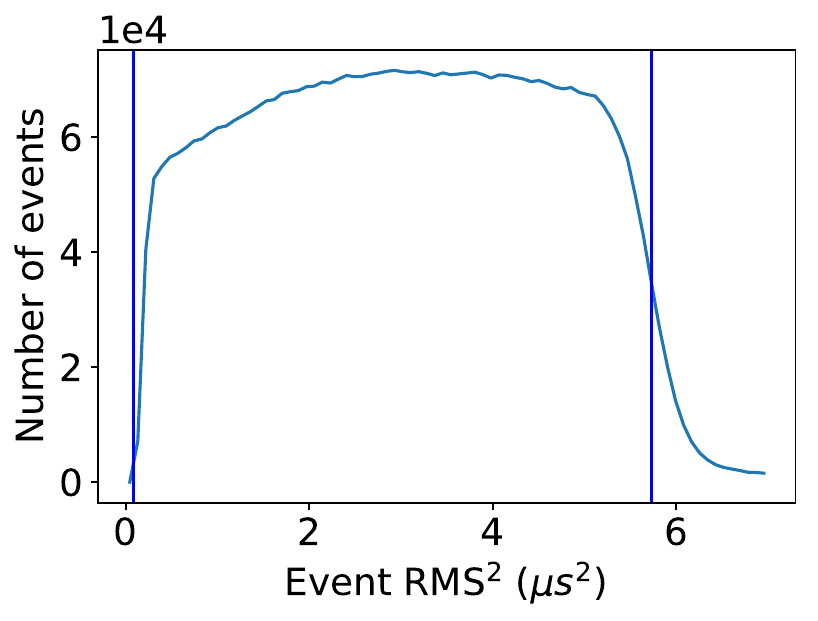}
  \centering
  \includegraphics[width=.50\linewidth]{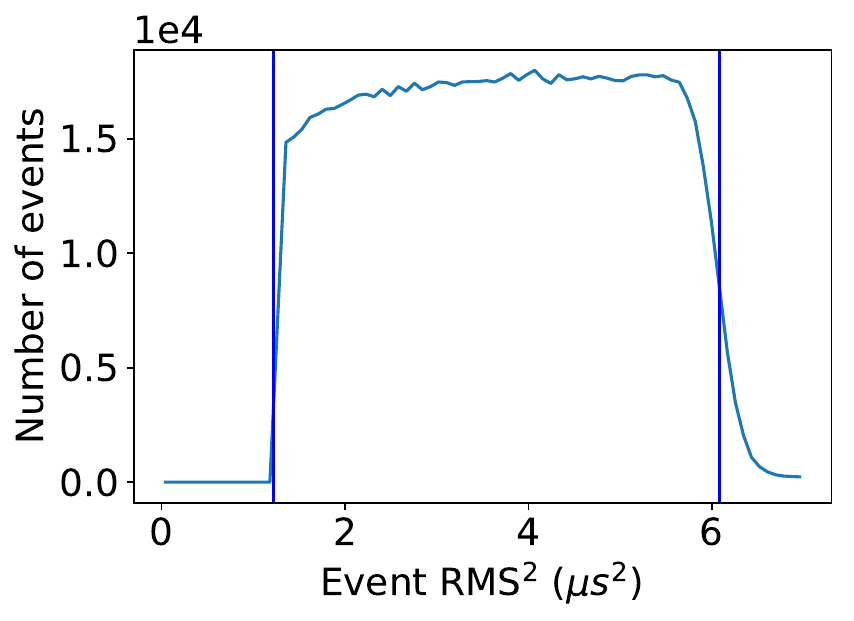}
\centering
\caption[Event z from width assignment without S1 reference using boundaries]{Distribution of mean squared (RMS$^2$) widths of $^{83\mathrm{m}}$Kr events, with boundary lines indicating corresponding minimum and maximum $z$ values of the detector as determined from the distributions as described in the text. Left: NEXT-White data. Right: NEXT-100 simulation.}
\label{fig:s1free_zhsits}
\end{figure}

It is notable that the  described method for assigning the positions of $^{83\mathrm{m}}$Kr events via diffusion did still rely on the use of S1 information indirectly, in order to build the calibration distributions of RMS$^2$ as a function of $z_{\mathrm{S1}}$.  In a detector which would not have the S1 information, this would not be possible.  Thus an alternative method must be used to calibrate the conversion between pulse width and $z$.  Because RMS$^2$ varies linearly with $z$ position, a known RMS$^2$ value at $z = 0$~mm and the maximal drift distance $z = z_{\mathrm{max}}$ is sufficient accomplish this tuning.  The distribution of observed values for RMS$^2$ shows a sharp rising edge for small values, but a long falling tail at maximal diffusion.  Nevertheless, in both simulation and data it was found that $z_{\textrm{max}}$ corresponds closely to the right half-max of the RMS$^2$ distribution. This is shown for NEXT-White data and NEXT-100 Monte Carlo in Fig.~\ref{fig:s1free_zhsits}. That the same method works for both data and simulation indicates that this ``boundary method'' is a reasonable and robust way of establishing the mean diffused pulse widths corresponding to the detector extrema without the need for S1-based tuning.%, as the calibration function is determined from the boundaries of the RMS$^2$ distribution.

\begin{figure}[htb!]
\centering
  \centering
  \includegraphics[width=.49\linewidth]{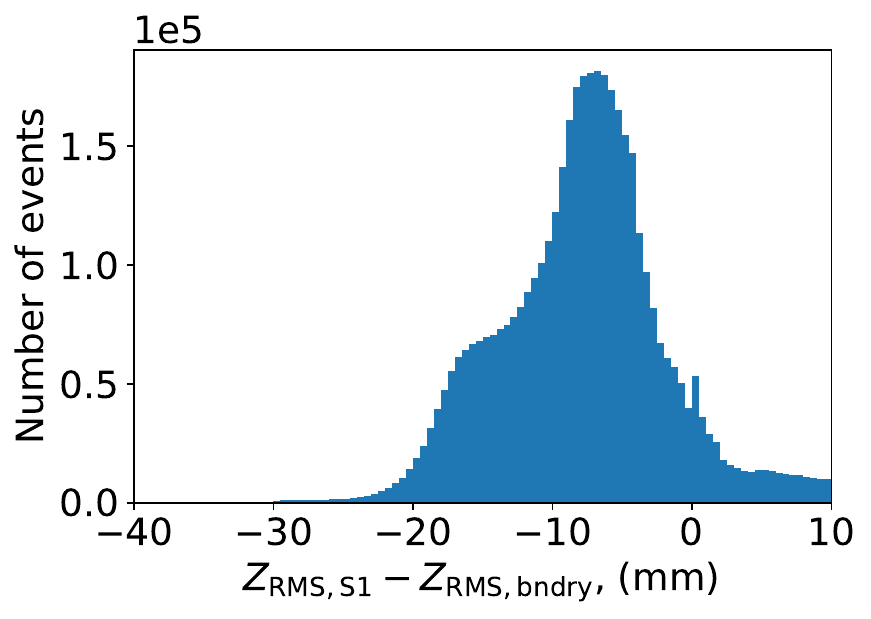}
  \centering
  \includegraphics[width=.49\linewidth]{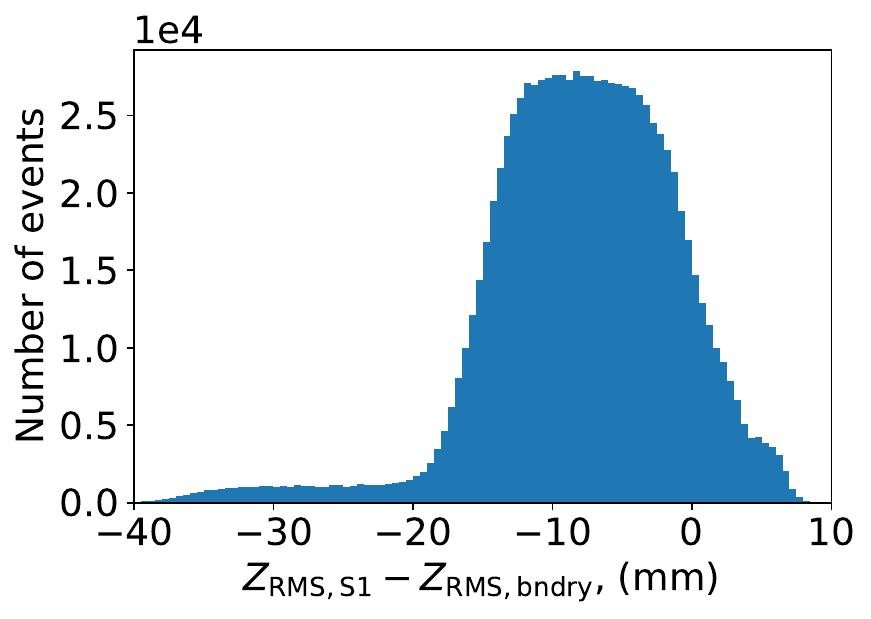}
\centering
\caption[Event z from width assignment without S1 reference for NEXT-100]{Distribution of differences between position of $^{83\mathrm{m}}$Kr events as assigned using the linear correlation between RMS$^2$ as a function of $z$ from $S1$ ($z_{\mathrm{RMS,S1}})$ and as assigned purely referencing the cutoffs of the RMS$^2$ distribution and the known detector boundaries in $z$ ($z_\mathrm{{RMS,bndry}}$), as described in the text. Left: NEXT-White data. Right: NEXT-100 simulation.}
\label{fig:s1free_diffhists}
\end{figure}

Fig.~\ref{fig:s1free_diffhists} compares the $z$ position obtained with a diffusion curve calibrated using S1, and one calibrated using the boundary method for NEXT-White data and NEXT-100 simulation. The distributions have some qualitative  differences, although in both cases errors tend to be slightly negative, with the boundary method assigning events as being slightly deeper (higher $z$) than the S1-referenced method. Errors are generally less than $20~$mm in magnitude in either case. This level of imprecision is not expected to have any significant effect on key detector performance metrics, given the expectation of free electron lifetimes greater than 5~ms, which correspond to 4500~mm at the NEXT-White drift field.

\section{Reconstruction of the $z$ position of $E>1.5$~MeV radiogenic electrons using diffusion}
\label{sec:he}

Extraction of the $z$ position of higher energy radiogenic events via diffusion is a more complex task than for the point-like deposits of $^{83\mathrm{m}}$Kr. The events of interest, including  photoelectrons and Compton electrons from gamma rays, as well as the two-electron signatures of either neutrinoless or two-neutrino double beta decays, present long, tangled topologies. The precise shape of the track will depend on both its local 3D structure and upon diffusion and the electroluminescent region response time profile.  To extract the spread from diffusion while accounting for the structure of the track in 3D space thus requires an analysis of the whole topology rather than a direct quantification of the S2 pulse shapes.  To this end, a neural network based approach was developed.

Deep neural networks have been employed for NEXT topological event reconstruction to distinguish between one-electron signatures of background events and two-electron double beta decay signatures.  A method was first proven in \cite{Renner_2017}, and honed in \cite{Kekic_2021} to achieve substantial performance improvements in event classification and background rejection techniques beyond traditional apporaches.  Those works use the double escape peak of $^{208}$Tl with energy of 1.6 MeV, as a monoenergetic calibration line of two-electron events, and use a network trained on Monte Carlo events to select the two-electron ``signals'' over a one-electron ``backgrounds'' from the local Compton continua from various higher energy gamma-ray lines.  Performance of the network was assessed based on how well the calibration peak at 1.6 MeV was extracted from backgrounds. This metric mirrors the requirement of distinguishing $0\nu\beta\beta$ events from $^{214}$Bi Compton events and $^{208}$Tl photoelectrons around the Q-value for $0\nu\beta\beta$ at 2.4 MeV.

\begin{figure}[t]
\centering
\includegraphics[width=0.99\linewidth]{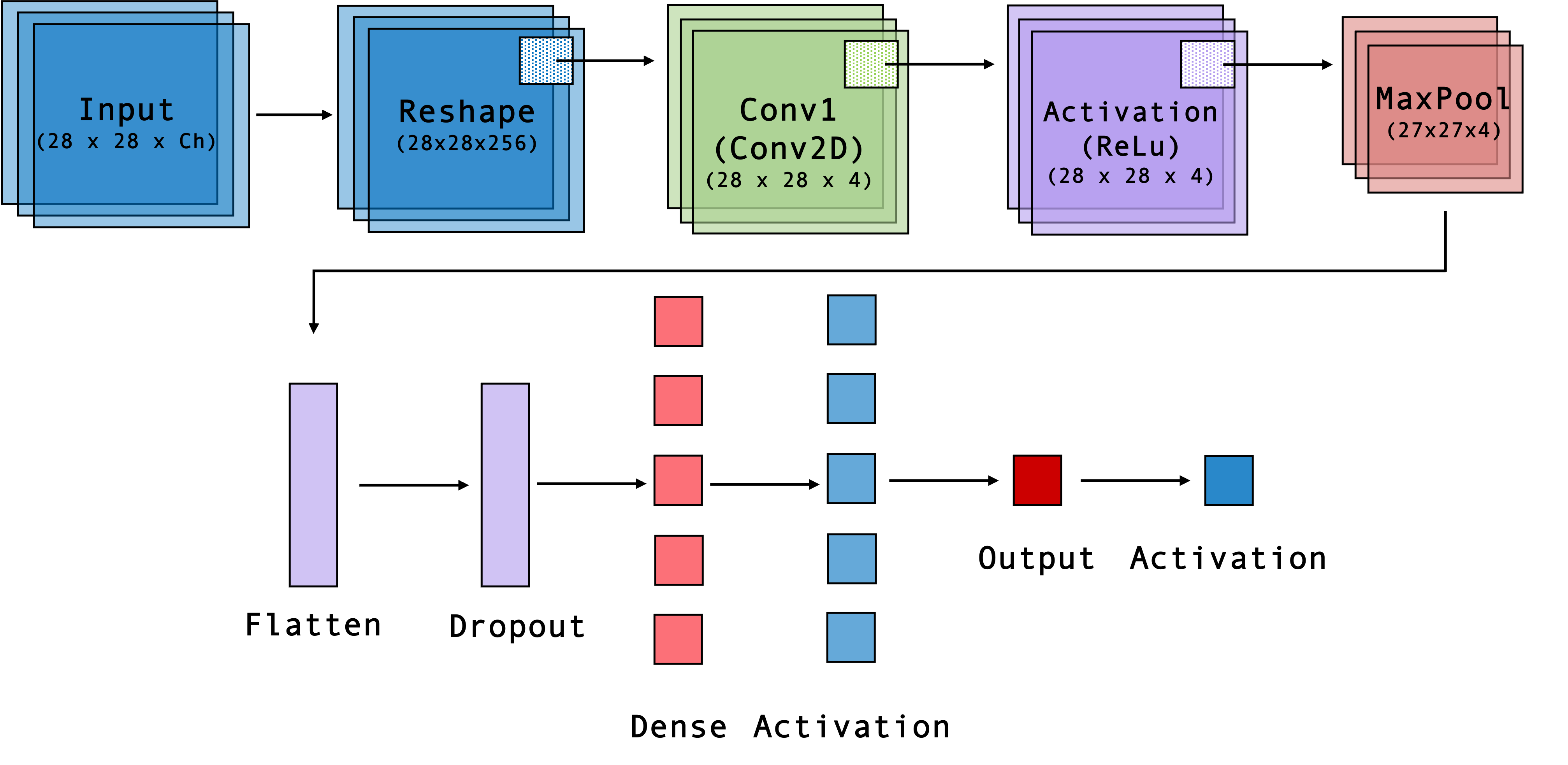}
\caption{Network architecture for $XY$ plane configuration. For all three planes, $XY$, $YZ$, and $XZ$ a sequential model is constructed. A permute layer is added to models $YZ$ and $XZ$ for dimensional order. Key features are extracted from layers in the top row, and classification of those features occurs with layers in the bottom row. }
\label{fig:NetArch}
\end{figure}

Making classification decisions about complex tracks using information about the full 3D image is a natural application for deep neural networks. For the application described here, however, a different network structure appeared optimal.  Whereas topology is a global decision about the track shape, the extraction of information on diffusion is a spatially localized process, and many local measurements may be expected to reinforce each other. This local information should be accessed while avoiding the possibility of over-training on complex track features. Thus the chosen network architecture is thus 1) convolutional, to measure features of local track regions; 2) shallow, to avoid encoding more than the simplest, local features into the classifier; and 3) trained with significant information dropout layers to avoid over-training on event topology details.

Both 2D and 3D convolutional approaches were assessed. Marginally better performance was achieved by utilizing three independent 2D convolutional networks. These networks are applied layer-by-layer to the event, and their outputs are combined in a single densely connected layer. Finally, the measured $z$ position, representing the barycenter of the event, is communicated to an output neuron. This improved performance of 2D over 3D networks is attributed to the larger number of extra free parameters in the 3D network, which ultimately provides slightly more of a training burden than an advantage given that the diffusion process acts essentially independently in each orthogonal direction.

\begin{figure}[t]
\centering
\includegraphics[width=0.99\linewidth]{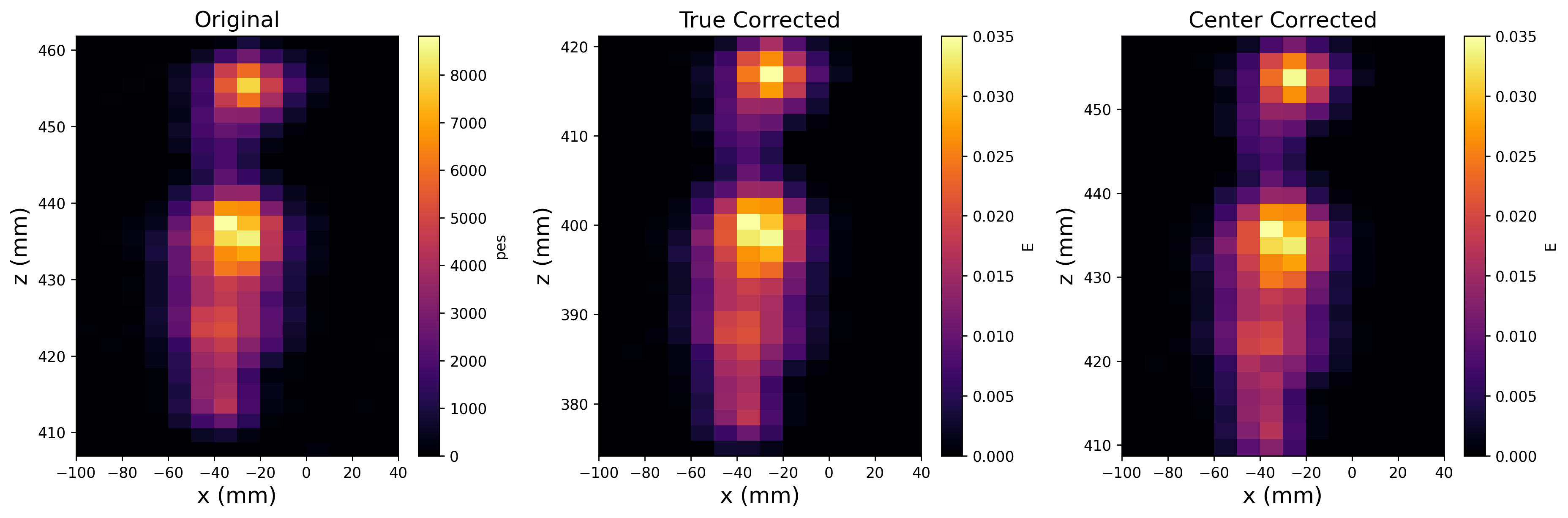}
\caption{Representation of the original uncalibrated event (left), the event after S1 calibration (middle), and the event post-diffusion calibration (right).}
\label{fig:EventExam}
\end{figure}

The three projections of the NEXT drift volume are inequivalent due to different event discretization scales in the transverse and longitudinal directions, with 10~mm SiPM pitch spacing transversely and 3~mm digitization distance longitudinally. There are also different longitudinal and transverse diffusion constants, and differing effects leading to event spreading during detection. Longitudinally the event is broadened by the EL crossing time of 2~$\mu$s whereas transversely it is spread by a non-Gaussian point-spread function associated with distribution of the VUV photons on the wavelength shifting plate.  For these reasons the network acting in the purely transverse $XY$ plane has different optimal parameters than the two acting in the longitudinal-transverse $XZ$ and $YZ$ planes, and they are trained independently.

The network architecture follows a sequential model where a series of layers are applied.  Each model is composed of a 2D convolutional layer along with an activation layer using the rectified linear unit function (relu), followed by a max pooling operation layer~\cite{chollet2015keras}. The model then uses a flatten and dropout layer to prevent over fitting during training. Two consecutive activation layers applying the relu function are accompanied by their own dense connected layer. The model is compiled for training using a mean of squares loss function between the true and predicted values. For each individual plane, a reshape and permute layer is incorporated before the 2D convolutional layer according to its $x$, $y$ and $z$ input dimension. A visual representation of the $XY$ network architecture is shown in Fig.~\ref{fig:NetArch}.

\begin{figure}[htb!]
\centering
    \includegraphics[width=0.9\textwidth]{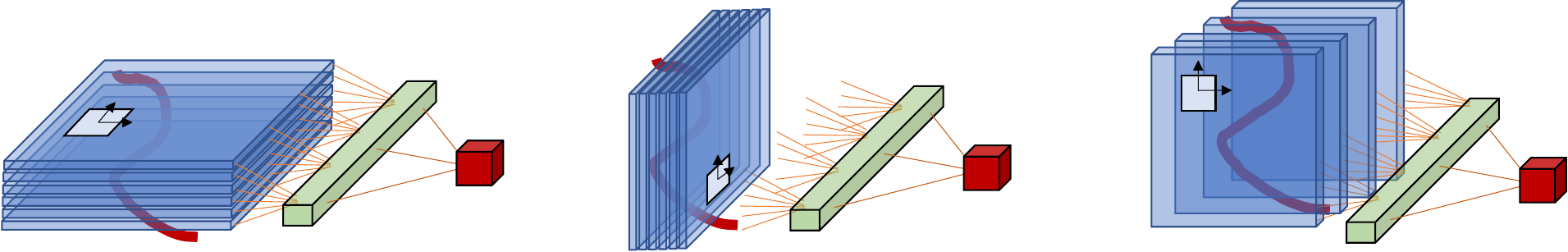}\hfill
    \begin{subfigure}[b]{0.32\textwidth}
    \centering
    \includegraphics[scale=.39]{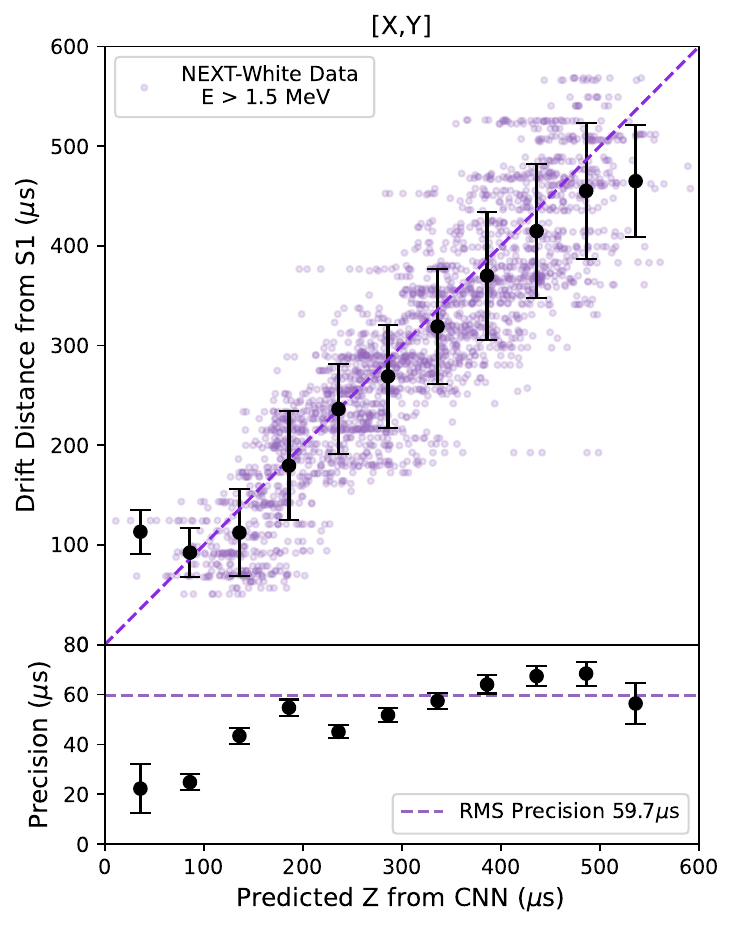}
\end{subfigure}
    \begin{subfigure}[b]{0.32\textwidth}
    \centering
    \includegraphics[scale=.39]{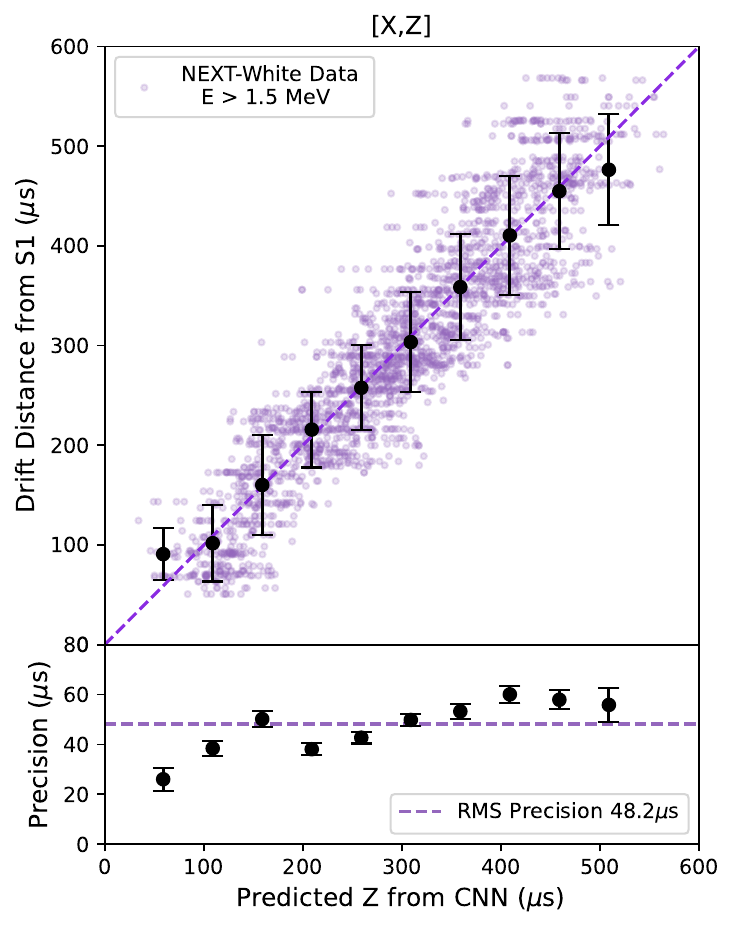}
\end{subfigure}
    \begin{subfigure}[b]{0.32\textwidth}
    \centering
    \includegraphics[scale=.39]{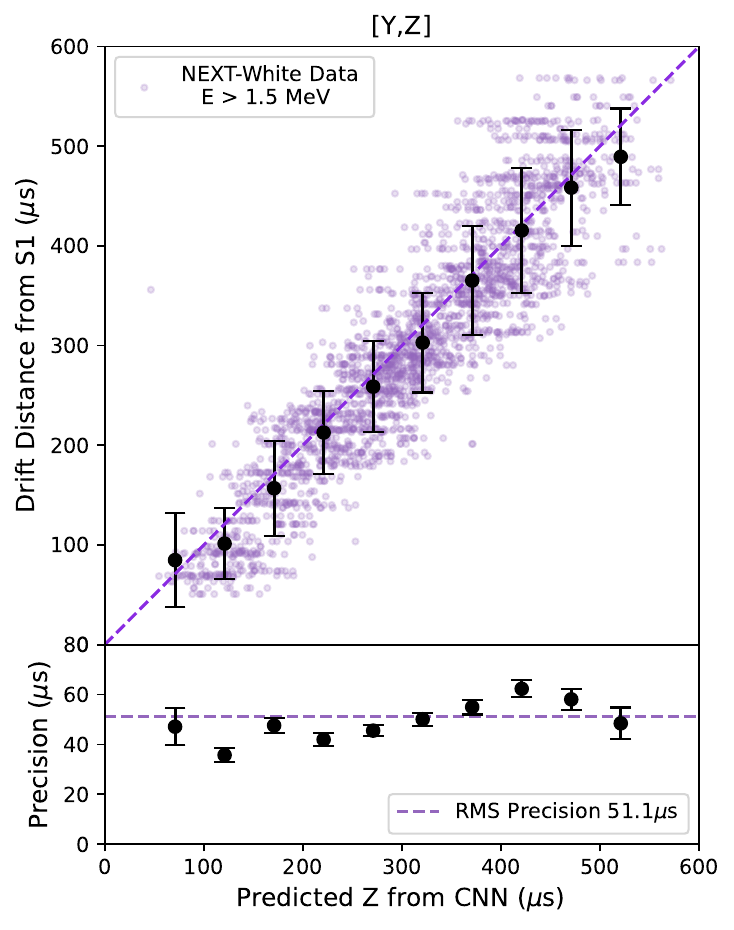}
\end{subfigure}
\caption{Top: Projection of 3 single-view (left: $XY$, middle: $XZ$, right: $YZ$) convolutional network output. 2D convolutional layers with filters concentrate of the effect of diffusion locally around track signature, a dense layer combines this information onto the output node, see Fig.~\ref{fig:NetArch} for more information. Bottom: The $z$ position can be obtained for long tracks using diffusion information from 3 plane configurations of the detector. The purple points represent the barycenter of NEXT-White data events that have passed selected cuts, the black points represent the mean for each binned slice with error bars that denote the standard deviation. The lower panels show precision (size of the standard deviation) with the RMS precision indicated by the dashed horizontal line.}
\label{fig:3 Views}
\end{figure}

The network is trained using real NEXT-White tracks containing one S1 pulse and one S2 pulse.  To produce S1-stripped data, raw events were artificially moved to have their mean $z$ positions at the center of the fiducial volume.  The raw hit charges were calibrated with krypton maps derived from the diffusion-only measurements of Sec.~\ref{sec:kr}, and applied as if the event was at the median $z$ position rather than its S1-reconstructed $z$ location.  This leads to a partially calibrated ``center-corrected'' event, which has an approximately reconstructed energy that we denote its ``center energy''.  This procedure implements the two calibrations that are possible with no S1 information: 1) correction of purely $XY$-dependent effects such as differential SiPM response, EL or WLS plate non-uniformity, and 2) the small adjustment to the $Z$ shape of the event from the electron lifetime correction, with longer-drifting electrons within the event being slightly more attenuated than shorter-drifting ones.  The overall event attenuation correction that typically uses $z$ from S1 is not applied.   Orthogonal subsets of such events are used as both training and test samples. Fig.~\ref{fig:EventExam} shows an event before calibration, after S1 correction, and after being center-corrected.

A data driven approach was developed to train the network that extracts $z$ position information from center corrected data. The network was trained to learn the S1-derived $z$ position for each center corrected event.  The training is run for 30 epochs for 3 uninterrupted passes. Training and validation were performed with the subset of fiducialized events with reconstructed energy above $E\geq 1.5$ MeV. These longer, most tangled events are not only the most challenging to extract the diffusion scale from, but also of the most interest for NEXT analyses including $0\nu\beta\beta$ searches, double-escape peak calibration of the NEXT topological signature, and calibration of the detector energy resolution using for example the $^{208}$Tl 2.6~MeV photo-peak.

\begin{figure}[htb!]
\centering
    \begin{subfigure}[b]{0.49\textwidth}
    \centering
    \includegraphics[width=\textwidth]{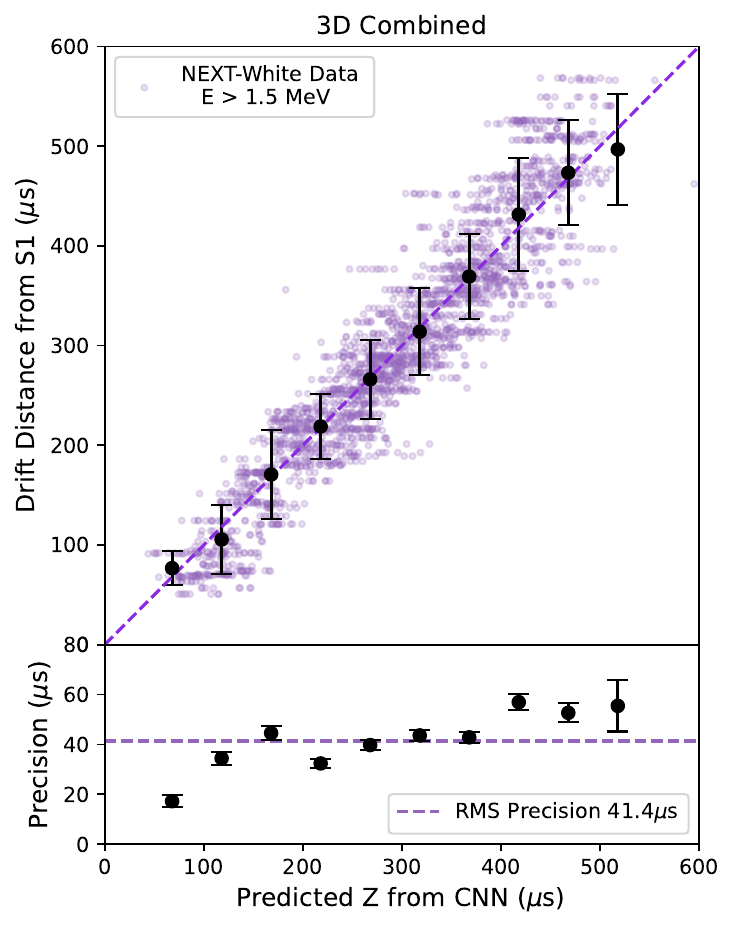}
\end{subfigure}
\hfill
    \begin{subfigure}[b]{0.49\textwidth}
    \centering
    \includegraphics[width=\textwidth]{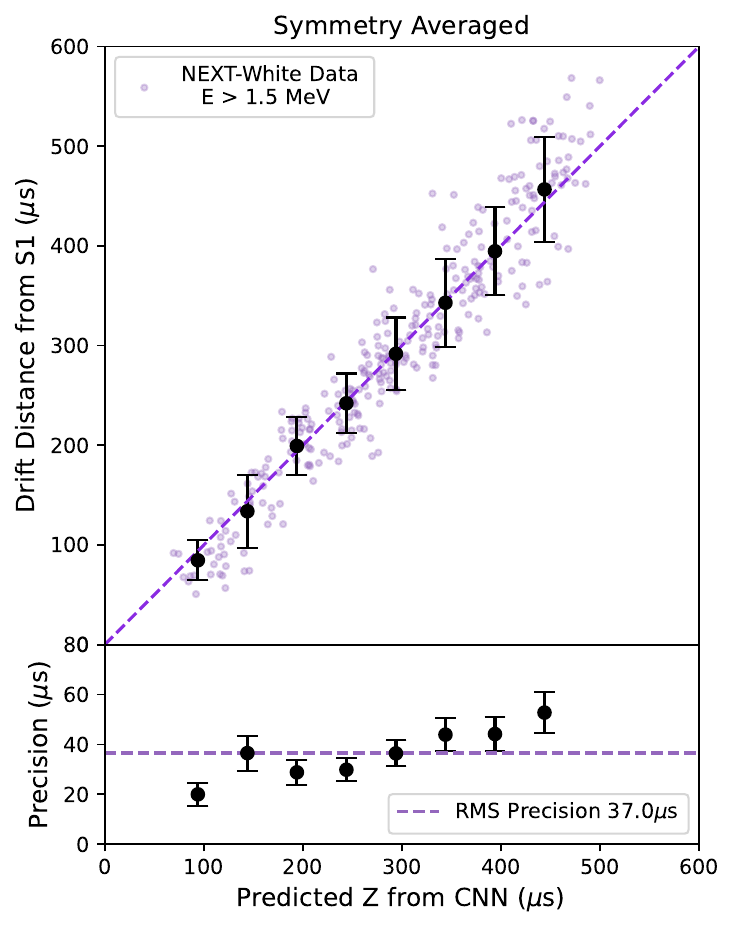}
\end{subfigure}
\caption{Predicted $z$ from CNN as a function of drift distance from S1. The purple points represent the barycenter of NEXT-White data events, the black points represent the mean for each binned slice along with error bars that denote the standard deviation on that bin. The lower panels show precision at each $z$ (size and error of the standard deviation per bin).  Left: The weighted sum of all 3 convolutional axes configurations.  Right: Also incorporating rotational symmetry by re-testing each event under each of its symmetry transformations.}
\label{fig:Data_Final}
\end{figure}

A total of 3,600 events passed these selection cuts. In order to maximize the statistical power of the training set each event was subjected to eight symmetry transformations in the transverse plane: this includes every combination of two possible mirroring operations and four rotations. The augmented training set is thus a factor of eight times larger than the original dataset, which improves the precision of the final network since its training is statistically limited.  This method can be used in the transverse plane but not in either of the longitudinal ones, since the front and back end of the track in the drift direction are not equivalent due to dissimilar diffusion scales at front and back.  This symmetrization of the network training is further exploited when placing events, averaging the result of operating the network on each of the eight symmetry transformations of each test event to provide the final estimate of its $z$ position.

\begin{figure}[htb!]
\centering
\includegraphics[width=0.8\linewidth]{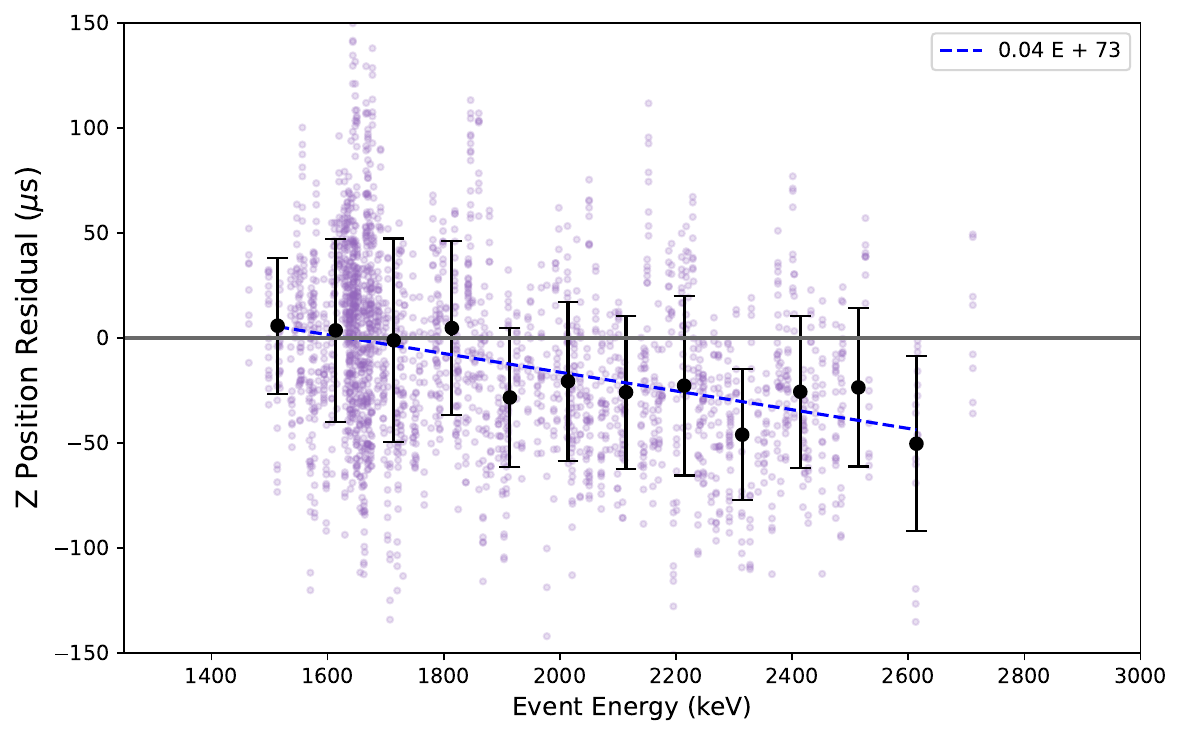}
\caption{Measurement of energy bias on $z$ residuals as a function of center energy. The best fit line in blue is used to correct events based on their center energy after initial $z$-placement by the neural network.}
\label{fig:energybias}
\end{figure}

The total energy for each event is normalized to a constant before either training or validation, so that the network is forced to extract the diffusion width from spatial information rather than using information on event brightness to estimate $z$. The location of the event is taken to be its barycenter, the summation of the hits $z$ times its energy divided by the events total energy.  After training, a small bias in the $z$ reconstruction as a function of energy was observed, and this is corrected with a linear function derived from the data, as shown in Fig.~\ref{fig:energybias}.  This correction is typically far smaller than the physical size of the event and makes only a marginal difference to the final average precision over the dataset.

The validation set is composed of 8\% of the data reserved from the training set to assess the network performance. The final prediction of the network was determined by an averaged sum of outputs from the $XY$, $YZ$, and $XZ$ networks. The average $z$-location precision from each plane for the data driven network are $XY$: 59.7~$\mu$s, $XZ$: 48.2~$\mu$s and $YZ$: 51.1~$\mu$s, as demonstrated in Fig.~\ref{fig:3 Views}; the precision on the averaged sum was improved to  41.4~$\mu$s. These can be converted into distance scales by multiplying by the drift velocity, approximately 0.91 mm/$\mu$s. A further small improvement was obtained by exploiting the symmetry properties of the detector, as described above.  The $z$ precision after this procedure was found to improved to 37.0~$\mu$s.  The final network performance is shown for the validation sample in Fig.~\ref{fig:Data_Final}. A linear relationship can be seen between the network predicted $z$ and drift distance from S1.

\begin{figure}[htb!]
\centering
\includegraphics[width=0.99\linewidth]{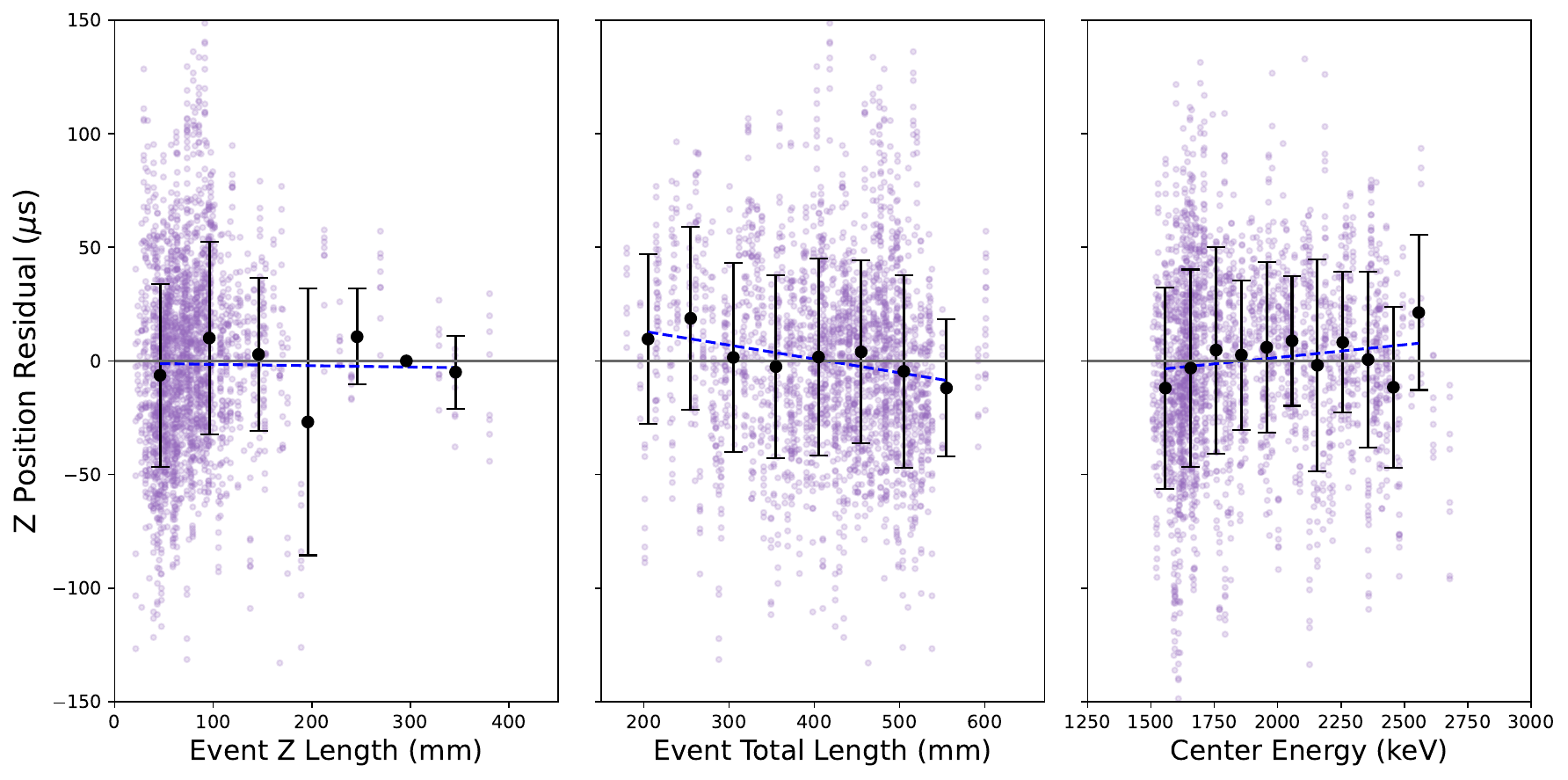}
\caption{Tests for  biases in the event $z$ precision as a function of event characteristic shape and energy: (left) $z$-extent, (middle) total event length, and (right) true event energy. The purple points indicate NEXT-White events after CNN application and center energy correction.}
\label{fig:3ZBiases}
\end{figure}

The performance of the diffusion-based $z$ reconstruction protocol was assessed as a function of event center energy, total length and length in the $z$ direction to test for possible biases. No strong dependencies of precision upon on these variables were observed.  These tests are shown in Fig.~\ref{fig:3ZBiases}.  

We thus conclude that events can be reconstructed in 3D space using diffusion to measure $z$, with a demonstrated precision of approximately 37 $\mu$s, or 33.6 mm.  This precision is small relative to the measured electron lifetime of between 5~ms and 14~ms in NEXT-White~\cite{next2022}, suggesting the method is sufficiently precise to calorimetrically correct event energies without $z$-positioning becoming a limiting factor for energy resolution in a diffusion-based reconstruction chain. Using the 7~ms lifetime of the run considered in this study, the implied energy resolution at Q$_{\beta\beta}$ would be modified from the 1\%~FWHM~\cite{NEXT:ecal2019} value measured in NEXT-White to 1.1\% FWHM, adding the uncertainty introduced by $z$ positioning in quadrature.  The method is also precise enough to reject false S1-S2 coincidences, with potentially improvements for signal selection efficiency or background rejection factors in future double beta decay analyses, and to fiducialize events to reject cathode-originated radiogenic backgrounds.

\section{Conclusions\label{sec:conc}}

In this article, several methods of determining event $z$ position using diffusion in a pure xenon time projection chamber were demonstrated. Fitting the pulse shapes of $^{83\mathrm{m}}$Kr S2 signals yields uncertainties that are generally less than $25 \textrm{ mm}$ in NEXT-White. Using Monte Carlo simulation, it is shown that this method can be extended to a larger detector such as NEXT-100, with similarly negligible degradation to energy resolution. This method was used to generate calibration maps using Kr that can be applied to events with energy >1.5~MeV, even in the case where S1 information is absent in their generation. 

A convolutional neural network based approach has been demonstrated to reconstruct the $z$ position for higher energy (E>1.5~MeV) events via diffusion. A data driven approach  was used to construct a training and validation set to derive the learned-S1 from the $z$ position. The final $z$ precision was found to be 33.6 mm, by averaging the weighted predictions of the  $XY$, $YZ$ and $XZ$ networks over symmetry configurations.  This is far smaller than the measured electron lifetime, suggesting promise as a method for longitudinal event reconstruction without S1.

This work thus provides a demonstration of a new way to identify $z$ positions in order to reject flawed events, and for allowing one to properly assign S1 and S2 peaks together even in cases where multiple events fall close together,  potentially offering higher event selection efficiencies and better background rejection capabilities. Furthermore, the results presented indicate the potential to use diffusion in a pure xenon time projection chamber to reconstruct the $z$ position of events even if no S1 signals are available.

\section*{Acknowledgements}
The NEXT Collaboration acknowledges support from the following agencies and institutions: the European Research Council (ERC) under Grant Agreement No. 951281-BOLD; the European Union’s Framework Programme for Research and Innovation Horizon 2020 (2014–2020) under Grant Agreement No. 957202-HIDDEN; the MCIN/AEI of Spain and ERDF A way of making Europe under grants PID2021-125475NB and the Severo Ochoa Program grant CEX2018-000867-S; the Generalitat Valenciana of Spain under grants PROMETEO/2021/087 and CIDEGENT/2019/049; the Department of Education of the Basque Government of Spain under the predoctoral training program non-doctoral research personnel; the Spanish la Caixa Foundation (ID 100010434) under fellowship code LCF/BQ/PI22/11910019; the Portuguese FCT under project UID/FIS/04559/2020 to fund the activities of LIBPhys-UC; the Israel Science Foundation (ISF) under grant 1223/21; the Pazy Foundation (Israel) under grants 310/22, 315/19 and 465; the US Department of Energy under contracts number DE-AC02-06CH11357 (Argonne National Laboratory), DE-AC02-07CH11359 (Fermi National Accelerator Laboratory), DE-FG02-13ER42020 (Texas A\&M), DE-SC0019054 (Texas Arlington) and DE-SC0019223 (Texas Arlington); the US National Science Foundation under award number NSF CHE 2004111; the Robert A Welch Foundation under award number Y-2031-20200401. Finally, we are grateful to the Laboratorio Subterr\'aneo de Canfranc for hosting and supporting the NEXT experiment.

\bibliographystyle{JHEP}
\bibliography{main}
% \bibliographystyle{JHEP}

% \providecommand{\href}[2]{#2}\begingroup\raggedright\begin{thebibliography}{10}

% \bibitem{Novella:2019cne}
% {\scshape NEXT} collaboration, P.~Novella et~al., \emph{{Radiogenic Backgrounds
%   in the NEXT Double Beta Decay Experiment}},
%   \href{http://dx.doi.org/10.1007/JHEP10(2019)051}{\emph{JHEP} {\bfseries 10}
%   (2019) 051}, [\href{https://arxiv.org/abs/1905.13625}{{\ttfamily
%   1905.13625}}].

% \end{thebibliography}\endgroup

% \bibliographystyle{JHEP}

\end{document}